\documentstyle[epsfig,12pt]{article}

\newcommand{\mycomm}[1]{\hfill\break
$\phantom{a}$\kern-3.5em{\tt===$>$ \bf #1}\hfill\break}
\newcommand{\mycommA}[1]{\hfill\break
$\phantom{a}$\kern-3.5em{\tt***$>$ \bf #1}\hfill\break}

\catcode`\@=11 
\let\rel@x=\relax
\newcount\timecount
\newcount\hours \newcount\minutes  \newcount\temp \newcount\pmhours
\hours = \time
\divide\hours by 60
\temp = \hours
\multiply\temp by 60
\minutes = \time
\advance\minutes by -\temp
\def\hour{\the\hours}
\def\minute{\ifnum\minutes<10 0\the\minutes
            \else\the\minutes\fi}
\def\clock{
\ifnum\hours=0 12:\minute\ AM
\else\ifnum\hours<12 \hour:\minute\ AM
       \else\ifnum\hours=12 12:\minute\ PM
            \else\ifnum\hours>12
                 \pmhours=\hours
                 \advance\pmhours by -12
                 \the\pmhours:\minute\ PM
                 \fi
            \fi
         \fi
\fi
}

\def\monthname{\rel@x\ifcase\month 0/\or January\or February\or
   March\or April\or May\or June\or July\or August\or September\or
   October\or November\or December\else\number\month/\fi}

\def\bold#1{\setbox0=\hbox{$#1$}     \kern-.025em\copy0\kern-\wd0
     \kern.05em\copy0\kern-\wd0
     \kern-.025em\raise.0433em\box0 }

\def\lsim{\mathrel{\mathpalette\@versim<}}
\def\gsim{\mathrel{\mathpalette\@versim>}}
\def\@versim#1#2{\vcenter{\offinterlineskip
        \ialign{$\m@th#1\hfil##\hfil$\crcr#2\crcr\sim\crcr } }}
\catcode`\@=12

\begin{document}
\def\beq{\begin{equation}}
\def\eeq{\end{equation}}
\def\MSbar {\hbox{$\overline{\hbox{MS}}\,$}}

\vskip 20pt

\begin{titlepage}
\begin{flushright}
{\footnotesize
CERN-TH-97/126\\
OSU~RN~327\\
SLAC-PUB-7566\\
TAUP-2434-97\\}
\end{flushright}
\begin{centering}
{\large{\bf
Pad\'e Approximants, Optimal Renormalization Scales,
and Momentum Flow in Feynman
Diagrams}}\footnote{
\baselineskip=14pt  Work supported in part by the Department
of Energy, contract DE--AC03--76SF00515.}\\
\vspace{.6cm}
{\bf Stanley J. Brodsky}\\
\vspace{.05in}
Stanford Linear Accelerator Center\\
Stanford University, Stanford, California 94309\\
e-mail: sjbth@slac.stanford.edu\\
\vspace{.4cm}
{\bf John Ellis}\\
\vspace{.05in}
Theoretical Physics Division, CERN, CH-1211 Geneva 23, Switzerland \\
e-mail: john.ellis@cern.ch \\
\vspace{0.4cm}
{\bf Einan Gardi} \,\,\, and \,\,\, {\bf Marek Karliner} \\
\vspace{.05in}
School of Physics and Astronomy
\\ Raymond and Beverly Sackler Faculty of Exact Sciences
\\ Tel-Aviv University, 69978 Tel-Aviv, Israel
\\ e-mail: gardi@post.tau.ac.il, marek@vm.tau.ac.il
\\
\vspace{0.4cm}
{\bf Mark A. Samuel}\\
\vspace{.05in}
Department of Physics, Oklahoma State University, \\
Stillwater, Oklahoma 74078, USA\\
e-mail: physmas@mvs.ucc.okstate.edu   \\ 
\vspace{0.6cm}
{\bf Abstract} \\
\vspace{0.35cm}
{\small
We show that the Pad\'e Approximant (PA) approach for resummation of 
perturbative series in QCD provides a systematic method for
approximating the flow of momentum in Feynman diagrams.
In the large-$\beta_0$ limit, diagonal PA's
generalize the Brodsky-Lepage-Mackenzie
(BLM) scale-setting method  to higher orders in
a renormalization scale- and scheme-invariant
manner, using multiple scales that represent
Neubert's concept of the distribution of momentum flow through a 
virtual gluon. If the distribution is non-negative, the
PA's have only real roots,
and approximate the distribution function 
by a sum of $\delta$-functions, whose locations and weights
are identical to the optimal choice provided by
the Gaussian quadrature method for numerical integration.
We show how the first few coefficients in
a perturbative series can set rigorous bounds on the all-order
momentum distribution function, if it is positive.
We illustrate the method with the vacuum 
polarization function 
and the Bjorken sum rule computed in the large-$\beta_0$ limit.
} 
\end{centering}

\end{titlepage}
\vfill\eject

\section{Introduction}

Pad\'e Approximants (PA's) 
are known to be useful in many physics applications,
including quantum field theory and statistical physics~\cite{padeworks}.
These applications of the PA method have recently been extended
to QCD~\cite{PA_QCD,Why}, where the method
has been shown to be
effective both in predicting unknown
higher-order coefficients and in summing the perturbative series.
In these applications of PA's to QCD,
the generic starting point is a 
perturbative series for some physical observable $A$ that has
been calculated exactly to some finite order
\begin{equation}
A\,\sim \,A_n\,\equiv \,x(C_0\,+\,C_1\,x\,+\,C_2\,x^2\,+\,\cdots\,+\,C_n\,x^n)
\label{ParSum}
\end{equation}
where the the expansion parameter $x$ is related to the renormalized
coupling constant by $x\equiv x(\mu
^2)=\alpha _s(\mu ^2)/(4\pi )$, where $\mu $ is the renormalization scale
in some scheme such as $\MSbar$. The corresponding
PA's are ratios of polynomials 
\begin{equation}
x[N/M]\,\equiv\, x\,\frac{p_0+p_1x+p_2x^2+...+p_Nx^N}{
1+q_1x+q_2x^2+...+q_Mx^M}  \label{PAdef}
\end{equation}
with $N+M=n$, chosen such that they reproduce
the known coefficients $C_0$ through $C_n$
when expanded back in a Taylor series. It is clear
that a PA (\ref{PAdef}) includes some higher-order effects: the hope is
that it
resums physically relevant higher-order contributions, so that $x[N/M]$
will be a better approximation to the observable $A$ than the original
truncated power series $A_{n}$.

In the absence of exact perturbative calculations for QCD observables
beyond three loops, the evidence for the effectiveness of the PA method
in QCD applications is mostly indirect:

\begin{itemize}
\item 
As already mentioned, PA's are successful in resumming the series as well
as in predicting higher-order coefficients in other models 
\cite{padeworks,PA_QCD}: see also \cite{KatStr} and a recent review
in \cite{Fischer}.
Among these,
an important example \cite{PA_QCD} is provided by the limit
of QCD where the number of light flavours $N_f$ becomes very large,
although there are important differences in the physics described by this
theory, due to its lack of asymptotic freedom.
The large-order renormalon behavior of the perturbative coefficients
is thought to resemble that of QCD, and there is a theorem that PA
predictions
for higher-order perturbative coefficients must converge if the
perturbative series is dominated by a renormalon.
\item  
Comparisons of the PA method~\cite{PA_QCD} with
other methods that seek to
optimize the perturbative result through a proper choice of scale
and scheme, such as the method of Effective Charges \cite{ECH}, 
the Principle of Minimal Sensitivity, \cite{PMS} and the
BLM scale-setting method \cite{BLM} show good numerical
agreement \cite{PA_QCD} for the Bjorken sum rule
and in certain cases also exhibit close algebraic relations~\cite{Why}.
\item
PA's were found \cite{PA_QCD} to reduce the undesirable
renormalization scale- and scheme-dependence of physical observables, as
compared to the partial sums on which they are based. 
Some understanding of this result was provided in~\cite{Why}, where it was
proven that diagonal $x[N-1/N]$ PA's become exactly scale invariant
when the $\beta$ function is approximated by its leading term.
This strongly
suggests that PA's resum correctly higher-order contributions 
associated with the running of the coupling constant: see also
\cite{MaxwellCR} for recent intriguing work along a somewhat
related approach.
\end{itemize}

Despite these pieces of evidence for the relevance of PA's for QCD, there
has so far, to our knowledge, been no direct diagrammatic 
interpretation of the summation by PA's. The magnitude of
this problem can perhaps be seen from the fact that a
Pad\'e function has simple poles in the coupling-constant
plane, and therefore cannot reproduce the expected factorial growth of the
perturbative coefficients. This
factorial growth is apparently an essential feature of
higher-order
contributions, appearing because of the flow of extremely
high or extremely low momentum through virtual gluon lines, ultraviolet
(UV)
and infrared (IR) renormalons, respectively, and also because of
the multiplicity of higher-order diagrams. 

It is clear from these considerations that PA's cannot account for the
full set of higher-order graphs. On the other hand, as reviewed above,
there are strong indications for the relevance of the
summation of higher orders of perturbative QCD by PA's. 
How can we interpret this summation in terms of Feynman diagrams?

Studies of higher-order perturbative QCD diagrams are often made
by first decomposing them in a skeleton expansion, in which
each term contains chains of vacuum polarization bubbles
inserted in virtual-gluon propagators. These have been studied in the BLM
approach, which seeks the optimal scale for evaluating each term
in the skeleton expansion.
The last step, the sum over skeleton graphs, is then
similar to summation of  perturbative contributions for a
corresponding theory
with
$\beta = 0$, i.e.,
a conformal theory~\cite{CSR,BGKL}.   We shall adopt a similar
procedure here.

In this paper, we consider a subset of graphs corresponding to a single
virtual-gluon exchange. 
We adopt the concept of the momentum
distribution function introduced by Neubert~\cite{mom}
(see also \cite{BB}): in the
large-$\beta_0$ limit,
 the all-order summation of diagrams is reduced to
a single integral over all scales of the running coupling constant, with
a weight function describing the distribution of momentum flowing
through the gluon line~\footnote{The relation of the optimal
renormalization scale and the BLM prescription to a weighted momentum
flow integral is also discussed in refs.~\cite{LM,CF}.}.
This is natural in QED calculations where the standard running
coupling
$\alpha(-k^2)$ sums all vacuum polarization corrections to the
photon
propagator. In QCD, the same feature is incorporated into the
$\alpha^V_s(-k^2)$
scheme defined from the potential for the color-singlet scattering of
two heavy colored test charges.
Since the coupling is
singular in the large-$\beta_0$ limit,  the integration over the gluon
momentum yields in general renormalon singularities.
We find that
PA's make a systematic approximation to the momentum distribution function,
thereby extending the leading-order BLM prescription.

In many physical cases, the
distribution function has been found empirically to be
non-negative~\footnote{In some cases
the UV cut-off must be chosen appropriately in order to achieve this.}.
If the distribution function is indeed non-negative, the resummed
amplitude
defines a so-called Hamburger function~\footnote{See~\cite{Baker} and
section 3.1 for the exact definition.}. Under this assumption, we 
obtain the following results:

\begin{itemize}
\item
One may
interpret the {\ }$x[N-1/N]$  PA as a discrete approximation to the
integral. This is because, for Hamburger functions, the poles
of the diagonal PA's are all real, corresponding to meaningful physical
scales, and their weights are all positive. 
\item
Moreover, there is a formal
connection between
PA's and the Gaussian quadrature method for numerical
integration~\cite{Baker}. The basis for this connection
is the observation that using
PA's amounts to approximating the distribution function by a sum of 
$\delta $ functions. The locations of these $\delta $ functions are
determined by the poles of the PA, and their weights are determined by the
residues. 
\item
Furthermore, one can use a known PA-based method to
bound rigorously the all-order distribution function, by using only the
first few coefficients of the perturbative series.
\end{itemize}

It is natural to ask in addition whether higher-order PA's converge to
the resummed
result, i.e., to the value of the integral over the exact,
continuous distribution function. This question is well posed
only when the integral itself is well defined. 
However, it is well known that this is not the case in the
presence of the IR
renormalons expected in QCD, and indeed we find that in general PA's do
not converge~\footnote{Although the PA predictions for the next 
individual terms in the perturbation series become progressively
more accurate.}. Only if there are no IR renormalons, meaning that
the momentum
distribution function completely vanishes for ``IR scales'' in addition to
being non-negative, in which case we would have a Stieltjes function
rather than just a
Hamburger function, would the integral be well defined on the positive
real axis in the $\alpha_s$ plane. In this case,
higher-order PA's do converge to the correct value.

The reader may find it useful if we relate our approach to that
of~\cite{BLM}. This is based on choosing the renormalization scale
of the coupling constant so that the next-to-leading (NLO) coefficient in
the leading $\beta_0$ series vanishes. The physical meaning of this
scale setting is that certain higher-order vacuum-polarization effects are
taken into account. The analysis of~\cite{mom} is a
generalization of the BLM~\cite{BLM} approach. The BLM choice of
scale amounts to approximating the continuous distribution function
$\rho(s)$ by a single $\delta$ function~\cite{mom}: 
$\rho(s) \longrightarrow C_0 \delta(s-s_{BLM})$.
As shown in \cite{Why}, in the
large-$\beta_0$ limit, the leading-order BLM procedure is also exactly
equivalent to a $x[0/1]$ PA, i.e., the BLM procedure is equivalent to a
single geometrical
series in the renormalized coupling constant at an arbitrary scale. Here we
show that higher-order diagonal $x[N-1/N]$ PA's in the large $ 
\beta_0$ limit can also be described naturally in the language of momentum
distribution functions: they correspond to approximating the distribution
function by $N$ $\delta$-functions, a multi-scale extension of the BLM
idea.
Several interesting extensions of the BLM method~\cite{BLM} have
been suggested in the past~\cite{GruKat,CSR,Rathsman}. The motivation was
to include higher-order effects, both within the large-$\beta_0$
approximation
and outside it. These suggestions were based on improving the leading-order
BLM scale~\cite{GruKat,CSR}, on introducing multiple scales to account
for non-leading $\beta_0$ terms~\cite{CSR}, 
improving the single BLM scale at higher order~\cite{BGKL},
and on setting the scheme by
tuning the coefficients of the $\beta$ function~\cite{Rathsman}. None of
these extensions, however, introduces multiple scales already {\em within}
the large-$\beta_0$ approximation, and in this sense these methods have no
natural description in the physically attractive language of momentum
distribution functions, as is done by diagonal PA's.

The outline of the paper is as follows: in the next section we review the
BLM approach and the concept of a momentum distribution function. In section
3 we study the relation between PA's and approximations to the momentum
distribution consisting of a sum of $\delta $ functions. We use some
known
mathematical results concerning Hamburger functions to draw the
above-mentioned conclusions
concerning the use of PA's in QCD. In section 4 we illustrate the 
application of our method to the vacuum polarization $D$ function
and the Bjorken sum rule
in the large-$\beta _0$ limit. Finally, in section 5 we 
give our conclusions and discuss some questions raised by this work.

\section{The BLM Method and the Momentum Distribution Function}

\subsection{Definition of the Momentum Distribution Function}

The BLM method~\cite{BLM} seeks to absorb most of the growth of the
higher-order perturbative QCD coefficients with an astute choice
of renormalization scale in lower-order expressions. Formally,
one arranges the perturbative
series for a generic QCD observable in a skeleton expansion,
whose coefficients are given by a conformal theory, and then seeks
an optimal scale to evaluate each term in the skeleton expansion.
The relation of this approach to the $x[0/1]$ PA in the large-$\beta_0$
approximation has been discussed in~\cite{Why}. In order to go
further, it is convenient to adopt the approach of~\cite{mom},
which considers the resummation of all orders of perturbative
corrections to physical observables that depend on a single external
momentum $Q^2$. The resummation method described takes into
account only the subset of graphs that can be 
described as an exchange of one effective virtual gluon,
which is the simplest example of a term in the skeleton
expansion. Resummation is achieved by using the running
coupling constant $\alpha _s(-k^2)$ at the vertices, where $k$ is the
momentum flowing through the virtual gluon, instead of a renormalized
coupling $\alpha _s(\mu ^2)$ at some fixed scale $\mu $. 

The result of
the improved calculation is denoted by $A_{res}$: 
\begin{equation}
A_{res}=\int d^4 k\,g(k,Q^2)\,\alpha_s(-k^2)=
\int_0^\infty dt\,w^R(t)\,x^R(tQ^2)
\label{A_res_R}
\end{equation}
where $g(k,Q^2)$ is the integrand of the Feynman diagrams. Clearly,
an infinite number of diagrams is taken into accounted by this 
integral. 
The function $w^R(t)$ in (\ref{A_res_R}) is interpreted as the
{\it momentum distribution function} characterizing the
virtuality of the exchanged gluon. It provides the
weighting in the integral over the running coupling 
$x^R(tQ^2)=\alpha_s^R(tQ^2)/(4\pi)$. The
superscript $R$ stands for the renormalization scheme, and serves to
remind us that both the distribution function and the running coupling
depend on the scheme, whilst the resummed result $A_{res}$ should be
{\it scheme-invariant}.

In practice, for physical examples, $A_{res}$ has been calculated 
to all orders 
only in the large-$\beta _0$ approximation, i.e., when 
the running of the coupling is controlled entirely by the 1-loop
$\beta$ function. Then 
\begin{equation}
x^R(tQ^2)=\frac{x^R(Q^2)}{1+\ln (t)\beta _0x^R(Q^2)}  
\label{running_x}
\end{equation}
where 
$\beta _0=\frac{11}3N_c-\frac 23N_f=11-\frac 23N_f$ for QCD.
All-order calculations in this approximation 
are possible because the
perturbative coefficients are proportional to the coefficients of the
large-$N_f$ limit \cite{BBB,LTM} in a non-Abelian
theory such as QCD, in which the higher-order
corrections are only due to fermion-loop insertions in the gluon
propagator. It has been argued~\cite{mom,LTM} that, since
$\beta _0$ is relatively large in QCD with a few
light flavors, this approximation should be quite good, and we
adopt it here.

We note that the integration in (\ref{A_res_R}) includes both high
momenta: $t\rightarrow \infty $ in the deep UV region and low
momenta: $t\rightarrow 0$ in the deep IR region. Whilst 
the UV integration region is well
defined, since renormalizability guarantees that $w^R(t)$ decreases fast
enough, the IR integration region is ill defined, due to the Landau pole
present in (\ref{running_x}).
This is how IR renormalons appear in this formulation, reminding us that
perturbation theory in QCD 
is not adequate to describe well 
long-distance physics with a truncated lowest-order  $\beta$ function.

\subsection{Renormalization-Scheme Independence}

In the large-$\beta _0$ approximation, the
dependence on the renormalization scheme can be removed 
from (\ref{A_res_R})~\cite{mom}. This is because the dependence
on the scheme enters only through the parameter ${\cal C}$, 
related to the finite part of a renormalized fermion-loop
insertion in the gluon propagator: in the $\MSbar$ scheme ${\cal
C}=-5/3$, and 
in the V-scheme ${\cal C}=0$. One observes that
$w^R(t)$ depends on ${\cal C}$, $Q^2$, and $\mu ^2$ only through the
combination $\mu ^2/(Q^2e^{{\cal C}})$, whilst $w^R(t)dt$ does not
depend on ${\cal C}$ at all. We follow~\cite{mom} in defining
\begin{equation}
\tau\,\equiv \,t \,\frac{\mu ^2}{Q^2 e^{{\cal C}}}
\label{deftau}
\end{equation}
and a scheme-invariant momentum distribution 
\begin{equation}
w(\tau): \,\,\, w(\tau)d\tau=w^R(t)dt
\label{invmeas}
\end{equation}
In this notation, (\ref{A_res_R}) becomes
\begin{equation}
A_{res}=\int_0^\infty w(\tau)x(\tau e^{{\cal C}} Q^2)d\tau  \label{A_res_w}
\label{newA_res}
\end{equation}
where $x(\tau e^{{\cal C}} Q^2)$ is scheme-invariant. Mathematically, the
parameter ${\cal C}$ plays here the same role as the scale $\mu$, and
therefore there is only one free parameter in the renormalization group.
Finally, we conclude that as long as we use the integral representation of
$A_{res}$, the choice of renormalization scheme does not change either 
$w(\tau)$ or the final result for $A_{res}$. We shall see shortly that this
is not true when one uses a finite-order Taylor series $A_n$ as an
approximation for $A_{res}$, where scale and scheme dependence do appear.
However, the use of diagonal $x[N-1/N]$ PA's eliminates the scheme and scale
dependence from the finite-order approximation for $A_{res}$~\cite{Why}.

For our purposes, it is convenient to rewrite (\ref{A_res_w}) again,
using $s \equiv \ln(\tau)$ as the integration variable: 
\begin{equation}
A_{res}=\int_{-\infty}^\infty \rho(s)x(e^{s+{\cal C}} Q^2)ds  \label{A_res}
\end{equation}
where $\rho(s)\,=\,w(\tau=e^s)\,e^s$. We then see from (\ref{running_x})
that
\begin{equation}
x(e^{s+{\cal C}} Q^2)\,= \,\frac{x^R(\mu^2)}{1+(s+{\cal C} 
-\ln(\mu^2/Q^2))\beta_0x^R(\mu^2)} \,= \,\frac{x^R(Q^2)}{1+(s+{\cal
  C})\beta_0x^R(Q^2)}
\end{equation}
and  
\begin{equation}
A_{res}=\int_{-\infty}^\infty \rho(s) \frac{x^R(\mu^2)}{1\,+\,\left(s+{\cal C 
}-\ln(\mu^2/Q^2)\right)\beta_0x^R(\mu^2)}ds  \label{A_res_mu}
\end{equation}
or 
\begin{equation}
A_{res}=\int_{-\infty}^\infty \rho(s) \frac{x^R(Q^2)}{1\,+\,\left(s+{\cal C} 
\right)\beta_0x^R(Q^2)}ds  \label{A_res_Q}
\end{equation}
We can think of $s$ as a scale parameter:
the long-distance physics is described
by $\rho(s)$ for negative $s$, while short-distance physics is described
by $\rho(s)$ for positive $s$.

\subsection{Perturbative Expansion}

We now turn to the perturbative treatment of $A_{res}$. 
We write the $n$-th order Taylor expansion for $A_{res}$ as 
\begin{equation}
A_{res}\,\sim \,A_n\,=\,x^R(\mu ^2)\sum_{k=0}^nC_k^R(\mu ^2)\left( \beta
_0x^R(\mu ^2)\right) ^k  \label{A_tay}
\end{equation}
where the coefficients $C_k^R(\mu ^2)$ are moments of the distribution
function $\rho (s)$: 
\begin{equation}
C_k^R(\mu ^2)\,=\,(-1)^k\,\int_{-\infty }^\infty \rho (s)\left( s+{\cal C} 
-\ln(\mu ^2/Q^2)\right) ^kds  \label{coef}
\end{equation}
The first coefficient $C_0$ is just the integral over the momentum
distribution function $\rho (s)$, i.e., its normalization, and does
not depend
on the scale and scheme. The second coefficient $C_1^R(\mu ^2)$ is related
to the average momentum flowing through the virtual gluon, and depends on
the scale and scheme through
\begin{equation}
C_1^R(\mu ^2)=C_1^R(Q^2)\,+\,C_0\ln(\mu ^2/Q^2)
\end{equation}
and 
\begin{equation}
C_1^R(\mu ^2)=C_1^V(Q^2)\,-\,C_0\left( {\cal C}-\ln(\mu ^2/Q^2)\right) 
\end{equation}
where the superscript $V$ stands for the $V$ scheme: ${\cal C}=0$, and
$R$ stands for a generic scheme characterized by 
some other value of ${\cal C}$. Higher-order
coefficients are given by higher moments of $\rho (s)$,
and depend on the scale and
scheme through higher powers of  $\left( {\cal C}-\ln(\mu ^2/Q^2)\right) $.

Consequently, at any finite order $n$ there is some residual 
dependence of the partial sum 
$A_n$ on the renormalization scheme and scale, through the combination 
$\left( {\cal C}-\ln(\mu ^2/Q^2)\right) $. This dependence is formally of the
next order in the coupling, but in practice it can be quite large and an
inappropriate choice of the scale and scheme can lead to misleading 
results, as
we show later for the example of the vacuum polarization $D$ function
- see  Fig.~1. 

\subsection{The BLM Prescription as a Narrow-Width Approximation}

As was shown in~\cite{Why} and mentioned in the
Introduction, the $x[N-1/N]$ PA based on $A_n :
n=2N-1$ eliminate the scale and scheme dependence completely from the
finite-order approximation to $A_{res}$.
As was also mentioned in the Introduction, the BLM prescription is based
on choosing a
renormalization scale so that the NLO coefficient 
vanishes in the large-$\beta_0$ approximation 
in which we work. We see from (\ref{coef}) that this 
translates into
\begin{equation}
\int_{-\infty }^\infty \rho (s)\left[s+{\cal C} -\ln(\mu_{BLM}^2/Q^2) 
\right]\,ds \,=\,0
\end{equation}
and therefore
\begin{equation}
(\mu _{BLM}^R)^2\,=\,Q^2 \exp\left( {\cal C}+\frac{\int_{-\infty }^\infty \rho
(s)sds}{\int_{-\infty }^\infty \rho (s)ds}\,\right) =\,Q^2e^{-C_1^R(Q^2)/C_0}
\label{mu_BLM}
\end{equation}
By construction in the BLM method, 
the NLO approximation $A_2$ to $A$ is
equal to the
leading-order approximation $A_1$. We obtain from
(\ref{A_tay}) and (\ref{mu_BLM})
\begin{equation}
A_{res}\sim A_1(BLM)=A_2(BLM)=C_0\,x^R\left( Q^2e^{-C_1^R(Q^2)/C_0}\right) 
\label{A_1_BLM}
\end{equation}
Note that the leading-order BLM result is scheme-invariant, 
whilst $\mu _{BLM}$ is scheme-dependent.

Another way to obtain the BLM result, which will be important for our
later generalization, is the following. Starting with (\ref{A_res}),
when only the first two coefficients $C_0$ and $C_1^R(Q^2)$ are known,
one can ask for what approximation to $\rho (s)$ is the
integral representation for $A_{res}$ equal to $A_1(BLM)$. 
The answer is~\cite{mom}
\begin{equation}
\rho (s)\,=\,C_0\,\delta \left( s+{\cal C}+\left( C_1^R(Q^2)/C_0\right)
\right) .  \label{rho_BLM}
\end{equation}
Note that this $\rho (s)$ is scheme-invariant, as it should be.
The BLM method can be thought of as an approximation to the
distribution function by a single $\delta$ function located at the BLM
scale. As stressed in~\cite{mom}, 
this is a good approximation only if the
distribution function is narrow. Information about the width 
and shape of the
distribution function are encoded in higher-order
coefficients, which are higher-order moments of the distribution function.

To see this in another way, we substitute (\ref{rho_BLM}) into
(\ref{A_res_Q}), obtaining 
\begin{equation}
A_{res}\sim A_1(BLM)=C_0\,\frac{x^R(Q^2)}{1\,-\,\left( C_1^R(Q^2)/C_0\right)
\beta _0x^R(Q^2)}.  \label{01PAa}
\end{equation}
which is another, exactly equivalent, representation of
(\ref{A_1_BLM}).
We stress again that the result is exactly scale- and
scheme-invariant. We could, just as well, write it as
\begin{equation}
A_{res}\sim A_1(BLM)=C_0\,\frac{x^R(\mu ^2)}{1\,-\,\left( C_1^R(\mu
^2)/C_0\right) \beta _0x^R(\mu ^2)}.  
\label{01PAb}
\end{equation}
In our notation, (\ref{01PAb}) is an $x[0/1]$ PA. The
leading-order BLM 
approximation in the large-$\beta _0$ approximation is therefore
equivalent to the
assumption that, when a given perturbative series is known up to NLO:
$x(C_0+C_1x)$,
an improvement is achieved by continuing it as a geometrical series
with the ratios of coefficients taken as $C_1/C_0$, obtaining:
$x(C_0+C_1x+{C_1}^2/C_0x^2+...)$. 
In the case of a positive distribution function $\rho (s)$, this is
probably a better approximation
than using the NLO truncated series at an arbitrary scale and scheme.
However, it is not a good approximation unless the distribution
function is narrow. In the next section, we will see how $x[N-1/N]$ PA's
provide better approximations at higher orders, whilst
maintaining the scale-independence property.

\section{Positive Momentum Distribution Functions and PA's}

\subsection{Positivity and Hamburger Functions}

A probabilistic interpretation of the momentum distribution function
has been proposed in~\cite{mom}: clearly,
an important issue for this interpretation is whether the distribution
function $\rho(s)$ is non-negative.
Moreover, if the sign of $\rho(s)$ changes, and in particular if
there
are large cancellations within the first moments, the BLM method may give
bad predictions. The BLM scale~(\ref{mu_BLM}) might
not describe accurately the typical virtuality of the gluon in such a
case.
The possibility of a change of sign in $\rho (s)$ is not excluded
in~\cite{mom}.
However, the examples given there suggest that the positivity of $\rho
(s)$ may
indeed be generic. This is certainly the case for the 
vacuum-polarization $D$ function: $\rho (s)\ge 0$ for any $s$,
and for the Bjorken sum rule, as we show in section 4.2.
In examples drawn from heavy-quark physics,
$\rho (s)$ becomes
negative only as an artefact of the UV cut-off, and it may well be that
$\rho (s)$ would be non-negative for an appropriate
choice of the UV regulator.

In the rest of this paper we
restrict our attention to cases where $\rho (s)\ge 0$ for any $s$.
This will be a crucial assumption for most of our results, which can be
checked only if $\rho (s)$ is calculated exactly, or if suitable
general theorems can be proven.

We begin by showing that the resummed result in the large-$\beta_0$ limit is
a Hamburger function of the renormalized coupling. 
Our starting point is expression (\ref{A_res_Q})
for the resummed result in the large $\beta _0$ approximation.
Since the integral is
scale- and scheme-invariant, we can work in the $V$ scheme where ${\cal
C}=0$,
and use the renormalization scale $\mu ^2=Q^2$, without loss of
generality.
In the large-$\beta _0$ approximation, it is convenient to define $ 
z=\beta _0x^V(Q^2)$ as the coupling constant. Thus we get: 
\begin{equation}
f(z)\equiv A_{res}/\beta _0=\int_{-\infty }^\infty \rho (s)\frac
z{1+sz}ds=\int_{-\infty }^\infty \frac z{1+sz}d\phi (s)  
\label{f}
\end{equation}
where $\phi(s)$ has the property that 
$\rho(s)=d\phi(s)/ds$. Then we define the moments
\begin{equation}
f_i \equiv (-1)^iC_i^V(Q^2)=\int_{-\infty }^\infty s^id\phi (s)
\label{f_i}
\end{equation}
for $i\geq 0$. 

If the moments are finite and $\rho (s)$ is a non-negative
function, i.e., $\phi (s)$ is a non-decreasing function, then
$ f(z)$ is a {\bf Hamburger function}~\cite{Baker}. 
Although $f(z)$ has a cut on the real
axis, it is well defined for complex arguments.
In the special case where the integration is limited to positive values of 
$s$, i.e.,  $\rho (s)=0$ for $s<0$, $f(z)$ is a {\bf Stieltjes function},
which is well defined on the positive real axis, since it has a cut
only for negative real $z$ values. In this case, the moments $f_i$ are all
positive and the perturbative
series $z\sum_{i=0}^\infty f_i(-z)^i$ is a Borel-summable asymptotic series.
However, QCD momentum distribution functions are not Stieltjes
functions, since long-distance effects make $\rho (s)$
non-zero for negative scales $s$.

At the present stage of knowledge of QCD, the statement that
$A_{res}$ should be a Hamburger function for a generic physical
observable is just an assumption, equivalent to assuming that $\rho(s)$ 
is positive. If
one knows only the first $n$ perturbative coefficients, one cannot
construct the exact
distribution and therefore cannot prove that it is indeed a Hamburger
function. However, Hamburger functions satisfy certain consistency
conditions: these can be checked already using the first few calculated
moments, and functions that do not satisfy them {\em cannot be}
Hamburger functions.
One such criterion is the
following: consider the determinants $D(m,n)$ defined by: 
\[
D(m,n)\,=\,\left| 
\begin{array}{cccc}
f_m & f_{m+1} & ... & f_{m+n} \\ 
f_{m+1} & f_{m+2} & ... & f_{m+n+1} \\ 
. & . &  & . \\ 
. & . &  & . \\ 
. & . &  & . \\ 
f_{m+n} & f_{m+n+1} & ... & f_{m+2n} \\ 
&  &  & 
\end{array}
\right| 
\]
In the case of a Hamburger function, the determinants $D(m,n)$
for even $m$ and any $n$ are positive~\cite{Baker}. Another
criterion is provided by the PA's discussed shortly: if the poles of
the $z[N-1/N]$ PA's include complex pairs, the momentum
distribution function cannot be a Hamburger function.

\subsection{Properties of PA's of Hamburger Functions}

We now use the
characteristics of Hamburger functions to draw conclusions on the use of
PA's for resumming the perturbative series for a positive
distribution function. 
We first draw attention to the theorem that a PA of a Hamburger
function has only real roots and positive weights. 
This means that successive PA's define approximations
to the distribution function in terms of
$\delta$ functions with the corresponding locations and weights.
The first such
approximation at NLO is provided by the $x[0/1]$ PA, which is equivalent
to the BLM method~\cite{Why}, and higher-order approximations are
provided by the $x[N-1/N]$ PA's. These
approximations to the momentum distribution function 
have the advantage of yielding scale- and scheme-invariant
results when one integrates, in the large-$\beta_0$ approximation.
As we discuss in more detail below, these approximations chosen by the PA
are identical to the optimal choices based on the all-order distribution
function according to the Gaussian quadrature method for numerical
integration. Then 
we show how, for a Hamburger function, the first few coefficients of the
perturbative series can be used to
set rigorous bounds on the all-order momentum
distribution function. Finally, we discuss the non-convergence of PA's due
to the presence of IR renormalons, and mention the non-physical case of a
Borel-summable series
that defines a Stieltjes function, for which PA's do converge.

In the language of momentum distribution functions, the 
perturbative coefficients may be interpreted as
moments of the {\em exact} distribution
function, as in (\ref{coef}). For a Hamburger function,
it can be shown~\cite{Baker} that the $z[N-1/N]$ PA
constructed from the partial sum $z\sum_{i=0}^nf_i(-z)^i$, 
where $n=2N-1$, may be written as: 
\begin{equation}
f(z)\,\sim \,z[N-1/N]\,=\,\sum_{i=1}^N\frac{r_iz}{1+q_iz}  
\label{diagPA}
\end{equation}
where (i) the locations of the poles $ {-}1/q_i$ are {\em on the real
axis},
and (ii) the weights $r_i$ are {\em positive}
for all $i=1,2,...N$.
\footnote{Note that 
the signs of the residues $r_i/q_i$ are determined by the 
signs of the $q_i$.}

\subsection{Approximations to the Momentum Distribution Function}

It is easy to construct approximations to $\rho (s)$ 
which yield (\ref{diagPA}) when substituted in 
(\ref{f}). These approximating distributions, which we
denote by $\rho _N(s)$, are given by 
\begin{equation}
\rho _N(s)=\sum_{i=1}^Nr_i\delta (s-q_i).   
\label{sum_delta}
\end{equation}
In words, the approximations are given 
by sums of $N$ $\delta$ functions, located at the points $q_i=-1/p_i$,
where the $z=p_i$ are the PA pole locations, and with weights 
$r_i$. Therefore PA's have very natural descriptions
as approximations to the momentum distribution function.
Correspondingly, $\phi (s)$, the indefinite integral of $\rho(s)$ is
approximated by monotonically non-decreasing piecewise-constant
functions with $N$ steps:
\begin{equation}
\phi _N(s)=\sum_{i=1}^Nr_i\theta (s-q_i)  
\label{phi}
\end{equation}
where $\theta(s)$ is the Heaviside function.

We note that the naive $n$th-order perturbation series has
no natural description in terms of the distribution function. Formally,
it may be expressed as a
singular function composed of the $n$ first derivatives of the Dirac
$\delta$ function: 
\begin{equation}
z\sum_{k=0}^nC_kz^k=\int_{-\infty }^\infty \widetilde{\rho }(s)\frac
z{1+sz}ds
\end{equation}
for $\tilde \rho (s)=\sum_{k=0}^n(C_k/k!)\delta ^{(k)}(s)$. It is clear that
the corresponding $\tilde \phi (s)$ is 
in general not a good approximation to $\phi(s)$: 
for one thing, it is not a monotonically non-decreasing function.

\subsection{Relation to the Gaussian Quadrature Method}

To underline further the utility of PA's, we note that
there is a formal mathematical relation between them and the Gaussian
quadrature method for numerical integration, which may provide
an opportunity to extend the resummation to include non-leading
terms in $\beta_0$.

The basic quadrature problem is a generalization of ours, 
namely to find a formula for the numerical
integration of a given arbitrary function $y(s)$ with respect to a positive
weight function $\rho (s)$: 
\begin{equation}
\int_a^by(s)\rho (s)ds=\sum_{i=1}^Ny(s_i)w_i+e_N
\label{Gauss}
\end{equation}
where $e_N$ is the error. One seeks the sampling points $s_i$ and
weights $w_i$ which minimize the error $e_N$. The 
well-known $N$th-order Gaussian
Quadrature method reviewed in~\cite{Baker} is to
choose the $N$ sampling points and corresponding weights
such that {\em any polynomial} function of maximal order $n=2N-1$
substituted for $ 
y(s)$ will be integrated {\em exactly}. The error 
for any other smooth function will in general
be small, since it results only from the difference between the exact
$y(s)$ and its best weighted polynomial approximation. The condition 
that any polynomial of maximal order $n$ will be
integrated exactly is: 
\begin{equation}
\int_a^bs^k\rho (s)ds=\sum_{i=1}^N{s_i}^kw_i
\end{equation}
for $k=0,1,...n$. 

It is easy to see that the required sampling points and
weights can be obtained from the $x[N-1/N]$ PA of the following Hamburger
function: 
\begin{equation}
\int_a^b\frac 1{1+zs}\rho (s)ds=\sum_{i=1}^N\frac{w_i}{1+zs_i}+O(z^{2N})
\end{equation}
In other words, the Gaussian quadrature formula for numerical integration
is obtained by replacing
the exact continuous distribution function $\rho (s)$ by a weighted sum of 
$\delta$ functions, with locations and weights determined by the
$x[N-1/N]$ PA of the corresponding Hamburger function.

Specializing now to the physical QCD problem, we start with the
following general expression for $A_{res}$: 
\begin{equation}
A_{res}=\int_{-\infty }^\infty x(e^{s+{\cal C}}Q^2)\rho(s)ds=\int_{-\infty
}^\infty x^V(e^sQ^2)\rho(s)ds
\end{equation}
In order to relate our problem to the Gaussian quadrature integration
problem, we identify the function $y(s)$ over which one integrates
in (\ref{Gauss}) with the running
coupling constant: $y(s)=x^V(e^sQ^2)$, and the weight function 
$\rho(s)$ with the Hamburger
momentum distribution function. The infinite integration range is not
expected to cause any trouble, since $\rho(s)$, being renormalized and IR
finite, is expected to vanish for large positive and negative arguments.
Thus our physical resummation problem is very close to the 
Gaussian quadrature
integration problem described above. The one difference in our
physical problem is that the 
running coupling constant is integrated with respect to a
weight function that is not fully known. The only 
pieces of information we have about this function are its
first few moments $C_0$ through $C_n$. This is to be contrasted with 
the mathematical integration problem, where the
weight function $\rho (s)$ is known exactly, but we limit the numerical
calculation to a certain order, for other reasons. 
However, at any given order,
the choices of scales and weights furnished by the $x[N-1/N]$ PA of the
corresponding Hamburger function is identical with
the optimal choice, according to
the Gaussian quadrature formula based on the {\em  exact} weight function.
Specifically, we obtain 
\begin{equation}
A_{res} \sim \sum_{i=1}^N r_i \,x^V\left(e^{q_i}Q^2\right)   
\label{Gauss_A}
\end{equation}
where the locations $q_i$ and weights $r_i$ are computed 
from the 
$z[N-1/N]$ PA (\ref{diagPA}) 
of the leading order of the large-$\beta_0$ series. 
As already remarked, the usual BLM approach corresponds to $N=1$.
Since the running
coupling
cannot be described as a finite-order polynomial in the scale, 
but rather as an infinite-order series, 
the PA's cannot yield the full all-order
resummation, but they do yield `optimal'
approximations to it.

One can also use equation (\ref{Gauss_A}) to resum effects that are
non-leading in $\beta_0$ in the running of the coupling constant,
simply by using a higher-order formula for $x^V(Q^2)$, involving 
$\beta_1$, $\beta_2$ and so on. We note that this procedure for 
using PA's to resum the series is {\em different}
from simply constructing the PA's of the
partial sums that contain non-leading terms in 
$\beta_0$. It seems likely to give more precise results,
though this requires further study.

\subsection{Bounding the Momentum Distribution Function}

In practice, one wishes to use the PA method and 
the momentum distribution 
formalism for QCD observables for which we know only a few low-order
perturbative coefficients. As we have shown, PA's provide an
approximation for the
distribution function, and one then integrates over the coupling
constant with this approximate distribution function as a weight. 
Mathematically, this is analogous to a 
moment problem, namely the construction of a distribution
function from its moments. If we assume that the momentum distribution 
is non-negative, we obtain
the so-called Hamburger moment problem. We 
have already seen how the $x[N-1/N]$ PA's
are related to approximations to the integral of the momentum distribution
function involving
$N$ steps (\ref{phi}). In this subsection, we use some
further mathematical theorems related to the Hamburger 
moment problem~\cite{Baker} to show
how the all-order distribution function can be
bounded rigorously - assuming that it is non-negative -
by continuous upper and lower bounds that are based on PA's,
using only the first few coefficients of the perturbative series. 

We start with the definition (\ref{f}) of the Hamburger
function, and 
assume that the moments $f_i$ defined by (\ref{f_i})
are known for $i=0,1...2M$ ($M\ge1$).
As before, we consider the partial sum $z\sum_{i=0}^n f_i(-z)^i$:
$n=2M+1$, and construct the diagonal PA as usual: 
\beq
z[M/M+1]\,=\,z\frac{A^M(z)}{B^{M+1}(z)}
\label{pademm+1}
\eeq
where $A^M(z)$ and $B^{M+1}(z)$ are polynomials of orders $M$ and $M+1$,
respectively. 
In principle, we cannot
construct this PA if $f_{2M+1}$ is unknown. However, 
the value of $f_{2M+1}$ will eventually be eliminated from our
final results for the bounds on the distribution function, so knowing it is
actually {\em not} essential for our present purpose.

Following the method described in part II, section 3.2 of~\cite{Baker},
we construct another PA:
\beq
z[M/M]\,=\,z\frac{C^M(z)}{D^M(z)}
\label{pademm}
\eeq 
where $C^M(z)$ and $D^M(z)$ are both polynomials of order $M$. 
We note that $z[M/M]$ is 
an off-diagonal approximant, and is therefore
not invariant under scale and scheme transformations,
in contrast to the $z[M/M+1]$ PA~\footnote{However,
our final bounds {\em are} scheme and scale independent.}.
An important observation is that $z[M/M]$, 
like the $z[M/M+1]$ PA, has only real roots. 

Using the two PA's (\ref{pademm+1}, \ref{pademm}), we 
introduce an auxiliary variable $w$ and
define
\beq
g(z)\equiv D^M(z)B^{M+1}(w)-D^M(w)B^{M+1}(z)
\label{gzdef}
\eeq
and
\beq
h(z)\equiv C^M(z)B^{M+1}(w)-D^M(w)A^M(z)
\label{hzdef}
\eeq
It is helpful to think of $g(z)$ and $h(z)$ as polynomials in $z$,
with $w$-dependent coefficients. One can then develop some intuition
for several interesting results on the ratio $h(z)/g(z)$,
provided in ref.~\cite{Baker}.
We do not prove them here: rather, we summarize them briefly and refer the
interested reader to~\cite{Baker} for details: 
\begin{itemize} 
\item The quantity $zh(z)/g(z)$ is an approximation to the 
original Hamburger function
$f(z)$, with the property that: 
\beq
\left(f(z)/z\right)g(z)-h(z)=O\left(z^{2M+1} \right) 
\eeq 
\item 
The quantity $zh(z)/g(z)$ can be written in the form 
\beq
z\frac{h(z)}{g(z)}=\sum_{j=1}^{M+1}\frac{z\rho_j}{1+z\zeta_j} 
\label{zhg}
\eeq 
where
$\rho_j$ and $\zeta_j$ depend on $w$. In particular, $\zeta_k=-1/w$ for
some $k$.  
\item 
Mimicking the relationship (\ref{f}) between $f(z)$ and $\phi(s)$,
one can construct from (\ref{zhg})
a monotonically increasing integral
distribution function $\psi(s)$, 
\beq
\psi(s)=\sum_{j=1}^{M+1}\rho_j\theta(s-\zeta_j) \eeq 
that, because of
the property mentioned above, has a point of increase at $s=-1/w$. 
\item
One can then define 
\begin{eqnarray}
\psi_{-}(s)\equiv\lim_{w=(-1/s)^-}\psi(s)\\ \nonumber
\psi_{+}(s)\equiv\lim_{w=(-1/s)^+}\psi(s) \end{eqnarray} 
and it can be shown that, for any $s$, 
\beq \psi_{-}(s)\le\phi(s)\le\psi_{+}(s)
\label{bounds} 
\eeq 
\end{itemize} 
These results imply that the
assumption that the distribution function $\rho(s)$ is non-negative
allows one not only to extract some good approximations to the all-order
distribution from the first few coefficients, but also to evaluate the
errors, by the construction of rigorous bounds on the distribution.  The
bounds (\ref{bounds}), just like the $x[N-1/N]$ PA's are
renormalization-scale and scheme invariant.  This is a direct
consequence of the fact that $zh(z)/g(z)$ is a $z[M/M+1]$ PA for any
value of the parameter $w$.

As an illustration, we
present here the explicit results for the bounding distribution
function in the case where only three coefficient in the perturbative
series are known,
namely $M=1$ in the above general analysis. This is both the simplest case
where such bounds can
be formed, and the maximal order for which QCD observables have been fully
calculated up to now. Thus, we start with a series:
\beq
f(x)=x(C_0+C_1x+C_2x^2)
\eeq
and use the $x[1/1]$ and $x[1/2]$ PA's to obtain the following bounds:
\beq
\psi_{-}(s)=\frac{(C_0s+C_1)^2}{C_0s^2+2C_1s+C_2}
\,\theta\left(s+\frac{C_1s+C_2}{C_0s+C_1}\right)
\eeq
and 
\beq
\psi_{+}(s)=\psi_{-}(s)+\frac{C_0C_2-C_1^2}{C_0s^2+2C_1s+C_2}
\eeq
We see that the bounds approach each other
as the series resembles more closely a
geometrical series, as intuitively expected.

Finally, we note a certain complementarity between
this method of constructing bounds for
the distribution function, and the previous PA
estimates of the distribution function. This arises
because the bounds are based on the perturbative series 
$z\sum_{i=0}^n f_i(-z)^i$ evaluated to some {\em even} order $n=2M$,
whereas the $x[N-1/N]$ PA 
requires the knowledge of the perturbative coefficients to some {\em odd}
order, $n=2N-1$. 

\subsection{Convergence of Higher-Order PA's}

Knowing that PA's resum a generic
perturbative QCD series, one may naively expect that
they should converge at high orders. However, it is
well known that perturbative series in QCD do not 
contain full information about
long-distance effects. 
In particular, IR renormalons appear in the formalism of
the momentum distribution function through 
the combination of the non-vanishing of the
distribution function $\rho(s)$ at negative $s$ with the 
Landau pole in the running of the coupling constant seen
in (\ref{A_res_mu}).
Mathematically, this implies that $A_{res}$ has a cut on the real
axis. From the physical point of view, this can only be avoided 
\cite{mom,DMW}
by postulating freezing of the coupling constant 
\cite{freezing}: for a recent discussion, see \cite{CaSt}.

From a mathematical point of view, 
it is interesting to look at another limit in which $A_{res}$ is well
defined on the positive real axis, namely when $\rho(s)$
vanishes for negative $s$.
If one assumes that $\rho(s)=0$ for negative $s$ in (\ref{f}), 
in addition
to the assumption that $\rho(s)$ is non-negative,
as already mentioned
one obtains a Stieltjes function:
\begin{equation}
f(z)= \int_0^{\infty} \rho(s) \frac{z}{1+sz} ds
\label{Stieltjes}
\end{equation}
rather than a Hamburger function. In this case,
$f(z)$ is well defined on the positive real axis, though it still
has a cut on the negative real axis, and therefore the formal Taylor
expansion of $f(z)\sim z\sum_{i=0}^{\infty} f_i(-z)^i$ has a zero radius of 
convergence. Unlike the Hamburger series, where the coefficients
have no definite sign, here the series oscillates in sign,
since the moments $f_i$ are all positive. 
There is a theorem~\cite{Baker} for Stieltjes series
that higher-order PA's {\em converge} to the {\em true Borel sum} of the
series, even though
the perturbative power series diverges.

In QCD, both the UV and the IR parts of the momentum distribution function
are expected to exist,
so we do not have the case of a Stieltjes series. 
Moreover, studies of different examples indicate
that the errors of PA's are particularly
large when the IR and UV parts are about equally important. 
However, we should like to stress that the
non-convergence of increasing-order PA's is expected to be 
much softer than that of the corresponding partial sums
in perturbative QCD. Whilst PA's
oscillate around the Cauchy Principal Value of the 
Borel-resummation integral~\cite{PA_QCD}, 
the partial sums diverge badly due to the zero radius
of convergence of the series. Thus the effects as well as the reasons
for the divergences are different.
 
\section{Some Worked Examples}

\subsection{The Vacuum Polarization $D$ Function}

As a first example of the general results of Sec.~3, we
choose the particular case of the vacuum-polarization $D$ function 
\beq
D(Q^2)=4\pi^2Q^2\frac{d\Pi(Q^2)}{dQ^2}=
N_c\,\sum_f Q^2_f \left[ 1+A_D(Q^2) \right ]
\eeq
to test our method. We neglect here a small
light-by-light contribution. For this example, the
all-order resummed result is known in the large-$\beta_0$
approximation~\cite{Broadhurst,BBB,LTM}: see also \cite{MaxwellPlanarD}.
Following~\cite{mom}, we write the momentum distribution function
in the form
\begin{eqnarray}
&\rho_D(s)&=
\,8C_fe^s\times\\\nonumber
&&\left\{\left(\frac{7}{4}-s\right)e^s+(1+e^s)
\left[L_2(-e^s)+s\ln(1+e^s)\right]
\right\}\,\,\,\,\,\,\,\,\,\,\,\,\,\,\,\,\,\,\,\,
\,\,\,\,\,\,\,\,\,\,\,\,\,\,\,\,s<0\\ \nonumber
&&\left\{1+s+\frac{e^{-s}}{2}\left(\frac{3}{2}+s\right)+(1+e^s)
\left[L_2(-e^{-s})-s\ln(1+e^{-s})\right]\right\}  
\,\,\,\,s>0\\ \nonumber
\label{rho_D}
\end{eqnarray}
where $L_2(x)\equiv-\int_0^{\infty}\frac{dy}{y}\ln(1-y)$ and
$\rho_D(s)$ is the weight function for the resummation integral:
\beq
A_D(Q^2)=\int_{-\infty}^{\infty}\rho_D(s)
\frac{x^R(Q^2)}{1+(s+{\cal C})\beta_0x^R(Q^2)}  
\label{A_D}
\eeq 
where we choose the renormalization scale $\mu^2=Q^2$, but still
allow for an arbitrary scheme $R$. 
Using equation (\ref{A_D}) we can obtain any required  
coefficient $d_k^R(Q^2)$:
\beq
d_i^R=(-1)^i \int_{-\infty}^{\infty}\rho_D(s)\left(s+{\cal C}\right)^i  
ds
\label{d_i}
\eeq
and the $n$th-order partial sum is: 
\beq
A_n(Q^2)=x^R(Q^2)\sum_{i=0}^n d_i^R(\beta_0x^R(Q^2))^i 
\label{partial}
\eeq
whose accuracy we now compare with PA's.

Fig.~\ref{D_val} presents the partial sums $A_n(Q^2)$ 
(\ref{partial}), as a function
of $n$ for $Q^2=2~GeV^2$. At this low value of $Q^2$ the
perturbative series starts to diverge already at 
relatively low order: $n\sim 5$, so the
differences between various renormalization schemes 
and other calculational approaches is readily
apparent. The horizontal continuous line 
in Fig.~\ref{D_val} represents the all-order
resummation of
the leading $\beta_0$ terms, as calculated by taking the Cauchy
Principal Value of the integral in (\ref{A_D}).
The two horizontal dash-dotted lines correspond to the maximal
uncertainty which is inherent to the integration in (\ref{A_D}), due
to the first IR renormalon.

Superposed on the all-order resummation results 
in Fig.~\ref{D_val}, we show the naive perturbation theory
partial sums in various schemes with $\mu^2=Q^2$. 
In this case, the $\MSbar$ partial sums 
of increasing order converge quite
nicely to the all-order result, whilst the the $V$-scheme results are
totally misleading. It is important, however, to realize this 
relative success of $\MSbar$ with $\mu^2=Q^2$ 
in this case has no known theoretical basis. 
There are other examples where $\MSbar$ with $\mu^2=Q^2$
is not a good choice, such as the Bjorken sum rule considered
in~\cite{PA_QCD}. It is clear that, if one chooses 
to evaluate partial sums, it is
essential to have some criterion for choosing 
an appropriate renormalization scale
and scheme~\footnote{As mentioned in section 2, 
the scale and scheme parameters play the same
mathematical role in the large-$\beta_0$ limit, 
so there is only one parameter to tune in this case.
Beyond the large-$\beta_0$ approximation, there are more
parameters to specify.}. 

An example of a judicious choice of scale 
is provided by the BLM criterion, which sets the scale such that 
the NLO contribution vanishes, as seen in
Fig.~\ref{D_val}, where the leading-order and NLO BLM results 
are indeed the same. The alternative and generalization to
higher orders that
we suggest, namely the $x[N-1/N]$ PA, does not have any scale ambiguity.
The $x[N-1/N]$ results presented in the figure are
on par with what one gets 
after {\em optimal} tuning of renormalization parameters 
within the usual schemes,
and certainly much better than 
the results with a generic choice of renormalization scheme and scale.

Next we consider the equivalence between taking the $x[N-1/N]$ PA of
the perturbative series and approximating the 
momentum distribution function by a weighted sum of $\delta$
functions, as discussed in section 3.3. The large-$\beta_0$ 
D-function momentum distribution of~(\ref{rho_D}) 
is indeed non-negative, and
the resummed result $A_D(Q^2)$ 
is therefore a Hamburger function. Hence the
results of section 3 are fully applicable to this physical example.
Since the final result in our method is scheme- and scale-independent,
we are free to choose the $V$-scheme with $\mu^2=Q^2$, which simplifies
the calculations. 
Starting with the $n$th-order perturbative series, where $n=2N-1$:
\beq
A_D \sim x^V(Q^2)\sum_{i=0}^n d_i^V(\beta_0x^R(Q^2))^i=\frac{1}{\beta_0}
\,z \sum_{i=0}^n d_i^V z^i
\eeq 
with the coefficients $d_i^V$ calculated from~(\ref{d_i}) with ${\cal
C}=0$, we construct the $z[N-1/N]$ PA. We determine
numerically the locations of its poles and their corresponding 
residues. As guaranteed by the general theorem~\cite{Baker},
all the PA poles $-1/q_i$ are real and 
all the weights $r_i$ are positive for any $i$.
We then find, for every $z[N-1/N]$ PA, the corresponding 
approximations to the momentum distribution function $\rho_D(s)$: 
\begin{equation}
\rho^D_N(s)=\sum_{i=1}^Nr_i\delta (s-q_i).   
\label{sum_delta_D}
\end{equation}
and to its indefinite integral $\phi_D(s)$
\begin{equation}
\phi^D_N(s)=\sum_{i=1}^Nr_i\theta (s-q_i) 
\label{phi_D}
\end{equation}
as described in section 3.

We illustrate in Fig.~\ref{D_delta} the way in which
the momentum distribution function $\rho(s)$ 
for the $D$ function is approximated
by a sum of $N$ $\delta$ functions, corresponding 
in panel (a) to the $x[N-1/N]$ PA's for
$N=1$ through $3$, and in panel (b) to the higher-order case $N = 12$. 
In each panel, the
exact continuous distribution $\rho_D(s)$ 
of~(\ref{rho_D}) is shown here as
a solid line. Superposed on it, we plot the locations 
$q_i$ and weights $r_i$ 
of the $\delta$ functions which compose 
the function $\rho^D_N(s)$ of~(\ref{sum_delta_D}).
At leading order, the $x[0/1]$ PA coincides with the BLM method, as
can be verified from the value of the scale parameter: $s_{BLM}^V \simeq
0.975$, which corresponds to $\mu_{BLM}^V \simeq 1.628\sqrt{Q^2}$.
The convergence of higher-order PA's is particularly clear in panel (b)
of Fig.~\ref{D_delta} - note that the vertical scale is logarithmic.

Fig.~\ref{D_theta} shows the 
corresponding integral 
$\phi(s)$ of the distribution function for
the $D$ function, including both the exact distribution $\phi_D(s)$
(continuous line) and the first three approximations
of~(\ref{phi_D}), that correspond
to the $x[N-1/N]$ PA's for $N=1,2,3$. We see how the $N$ steps imitate
the shape of the all-order momentum distribution,
based only on knowledge of the first few moments of the distribution.
   
Finally, in Fig.~\ref{D_bound} we illustrate 
the PA-based bounding technique of section 3.5,
for the same example of the vacuum-polarization $D$ function. The exact
integral distribution function $\phi_D(s)$ is again plotted as a
continuous line. We have
calculated the upper and lower bounds in the two
simplest cases, the first being based on 
the $x[1/2]$ and $x[1/1]$ PA's and
requiring knowledge of $d_0, d_1$ and $d_2$ (plotted as a 
dashed line), and the second being based on the $x[2/3]$ and $x[2/2]$
PA's and requiring the knowledge of $d_0$ through $d_4$ (dotted line).
As before, the calculation was done in the $V$ scheme. 
Just as for the $x[N-1/N]$ PA,
if one uses the bounds on the distribution function in order
to integrate over the coupling constant, one obtains a scale-
and scheme-independent result. 

We conclude this subsection with
an interesting empirical finding, for which we have no good
explanation. As can be seen in 	Fig.~\ref{D_bound},
the arithmetic mean of the upper and
lower bounds on the distribution function 
is quite close to the exact all-order result. This may very well
be a coincidence, but also might have some deeper 
theoretical justification.

\subsection{The Bjorken Sum Rule}

As an indication that the above example is not isolated, we now
consider the perturbative QCD series for the Bjorken sum rule, 
again in the large-$\beta_0$ approximation. This series is
known to have a particularly simple structure in the Borel
plane \cite{BroadhurstKataev,LTM}:
\begin{equation}
B(u) = {4 \over 9} {1 \over (1 +  u)}
- {1 \over 18} {1 \over (1 + {1 \over 2} u)}
+ {8 \over 9} {1 \over (1 -  u)}
- {5 \over 18} {1 \over (1 - {1 \over 2} u)}
\label{bjborel}
\end{equation}
where $B(u)$ is the Borel transform,
containing only four simple poles. 
Using an inverse Laplace integral \cite{mom} we can derive from
({\ref{bjborel}) the corresponding momentum
distribution function:
\begin{equation}
\rho (s) = 4\left[
\left( {8 \over 9} e^{{-s}} - {2 \over 9} e^{{-}2 s}\right) \theta (s)
+ \left({16 \over 9} e^{ s} - {10 \over 9} e^{2 s}\right) \theta ({-}s)
\right]
\label{bjmdf}
\end{equation}
This momentum distribution function is plotted in Fig.~5,
where we see explicitly that it is positive, and hence defines a Hamburger
function, as is the vacuum-polarization $D$ function in the
large-$\beta_0$ limit. 

It is clear that the steps carried out for
the previous example, namely the approximation of $\rho (s)$ (\ref{bjmdf})
by a sum of $\delta$ functions, the evaluation of the corresponding
integral $\phi (s)$ and the establishment of PA-based bounds, can
be carried out in a similar way, but we do not enter here into the
details.

\section{Conclusions}

This paper has been devoted to analyzing the PA
method in QCD, and to understanding the reasons for its success in
resumming perturbative series, which were
previously unclear~\cite{padeworks,PA_QCD}. 
We now understand that rigorous
conclusions can be drawn in the large-$\beta_0$ limit 
regarding the use of PA's. In particular, we find that
the $x[N-1/N]$ PA's are the most appropriate
for resumming the series, because of two important characteristics.
\begin{itemize}
\item
They are scale and scheme invariant \cite{Why}.
\item
If the momentum distribution of a virtual gluon in a generic
QCD observable is
non-negative, the resummation integral - which is a weighted average 
of the running coupling - defines a
Hamburger function. In this case, the 
\hbox{$x[N-1/N]$} PA is also the result of
an exact integration of the coupling constant over the $N$th-order optimal
approximation to the momentum distribution function. Therefore, the
resulting resummation 
makes full use of the first $n+1=2N$ coefficients of the series that
are known.
\end{itemize}
We also saw how the assumption of positivity of
the momentum distribution function
allows one to construct scale- and scheme-invariant 
bounds on the all-order
distribution, using knowledge of the first few perturbative
coefficients.

There are two main questions that our work raises. The first is: What
are the conditions for our conjecture on the positivity of the 
distribution function to be valid?
In some cases, where the large-$\beta_0$ all-order resummation
has been performed, this can be checked explicitly, as was done for
for the $D$ function in~\cite{mom} and for the Bjorken sum rule in this
paper. In other cases, where only a
few first coefficients are known, 
it would only be possible to disprove the conjecture,
as discussed in section 3.1. It would be very interesting to
find theoretical justification why the momentum distribution function
should be positive.

The second question is: How can one extend the PA's method, and use it
outside the large-$\beta_0$ limit?  It is obvious that the rigorous
results we have obtained
in this limit cannot be extended in a straightforward
manner.  The naive approach of using the
$x[N-1/N]$ PA of the full series is not very well motivated by the
momentum distribution concept, and gives a wrong functional 
dependence on the
number of colors and light flavors. An alternative, which is motivated
by the Gaussian quadrature integration procedure, is 
described in section 3.2. The basic idea is to use the scales and
weights of the corresponding large-$\beta_0$ series, and a
higher-order formula for the running coupling constant. This
method does not fully use the perturbative coefficients that are
known, so there is still room for further improvement.  

Despite the persistence of these open questions, we feel that
the analysis of this paper has contributed to a useful theoretical
foundation for the use of PA's in applications to perturbative QCD,
and also clarified in a useful way the relation of the BLM method to the
PA approach. We hope that this paper may serve as a helpful building block
in the search for an eventual optimal strategy for exploiting the
information contained in perturbative QCD series. Higher-order
QCD calculations have recently taken several impressive steps forward.
However, it seems in many cases unlikely that the following terms
in the series will be forthcoming in the foreseeable future. Moreover,
we shall in any case only have access to a finite set of exact terms
in any perturbative QCD series. Therefore tools to optimize the return
on the investment made in exact calculations will always be a welcome
input to making precision tests of QCD and understanding its place in an
eventual unified theory. We believe that both PA's and the BLM method
both contribute to the fashioning of these tools.

\vfill\eject
\begin{flushleft}
{\large\bf Acknowledgments}
\end{flushleft}

The research of E.G. and M.K.  was supported in part by the Israel
Science Foundation administered by the Israel Academy of Sciences and
Humanities, and by a Grant from the G.I.F., the German-Israeli
Foundation for Scientific Research and Development.
The research of M.A.S. was supported in part by the U.S.
Department of Energy under Grant No. DE-FG05-84ER40215.

\newpage
\begin{figure}[htb]
\begin{center}
\mbox{\kern-0.5cm
\epsfig{file=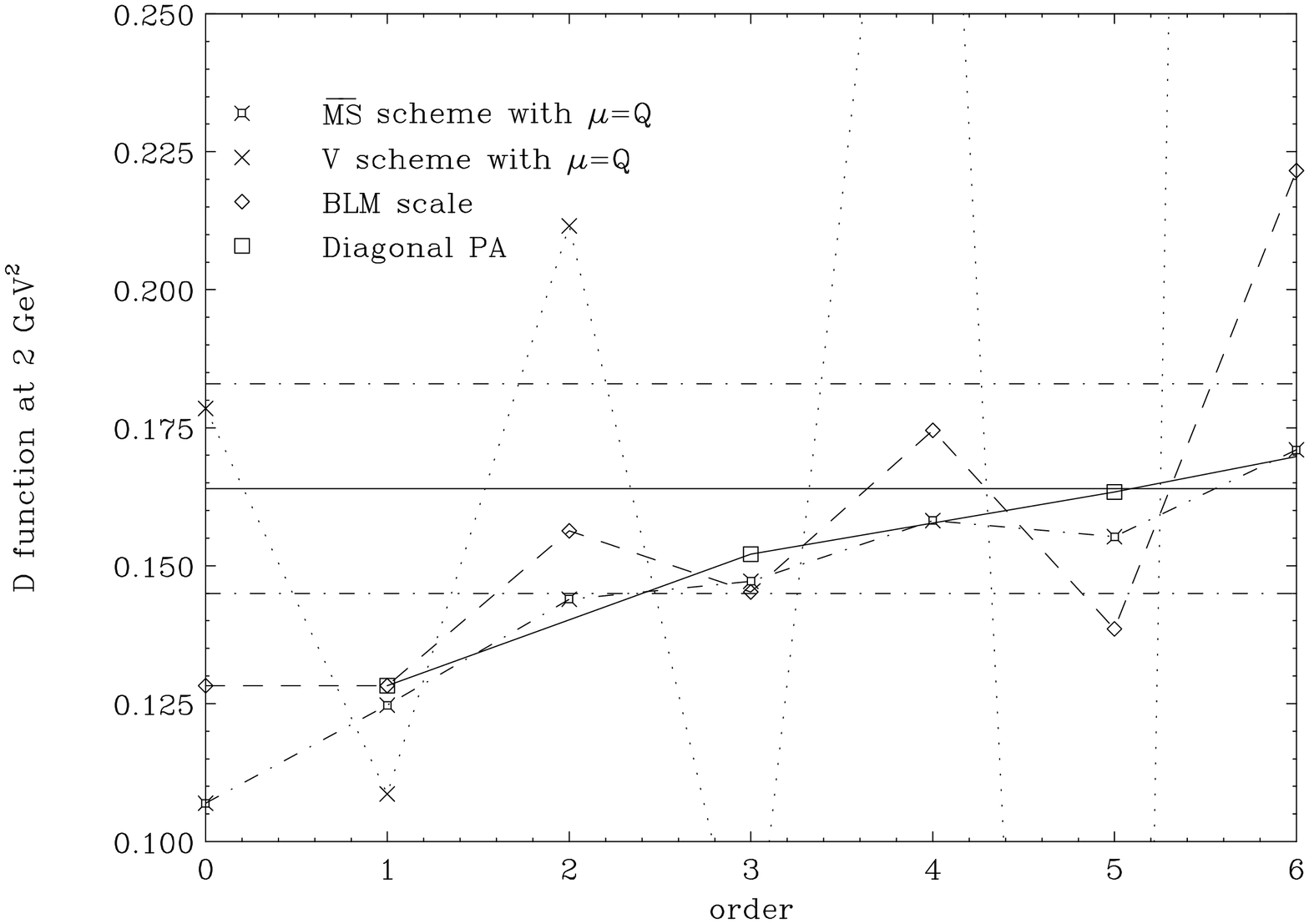,width=10.0truecm,angle=90}
}
\end{center}
\caption{Increasing-order results for the vacuum-polarization 
$D$ function in
the large-$\beta_0$ limit, as calculated in different renormalization
schemes/scales - $\MSbar$ with $\mu^2=Q^2$, the $V$ scheme with
$\mu^2=Q^2$, the BLM method, and diagonal PA's. The horizontal
continuous line
corresponds to the Cauchy Principal value of the Borel integral
of the perturbative series, and the
horizontal dash-dotted lines describe the maximal intrinsic ambiguity of this
integration due to the first IR renormalon pole. Note that the diagonal PA's
can be constructed starting with any definition of the coupling, i.e.,
using any scale and scheme, to give the results that are
plotted. Note also that leading-order result in the BLM method 
coincides exactly with the $x[0/1]$ PA. }
\label{D_val}
\end{figure}

\newpage

\begin{figure}[htb]
\begin{center}
\mbox{\kern-0.5cm
\epsfig{file=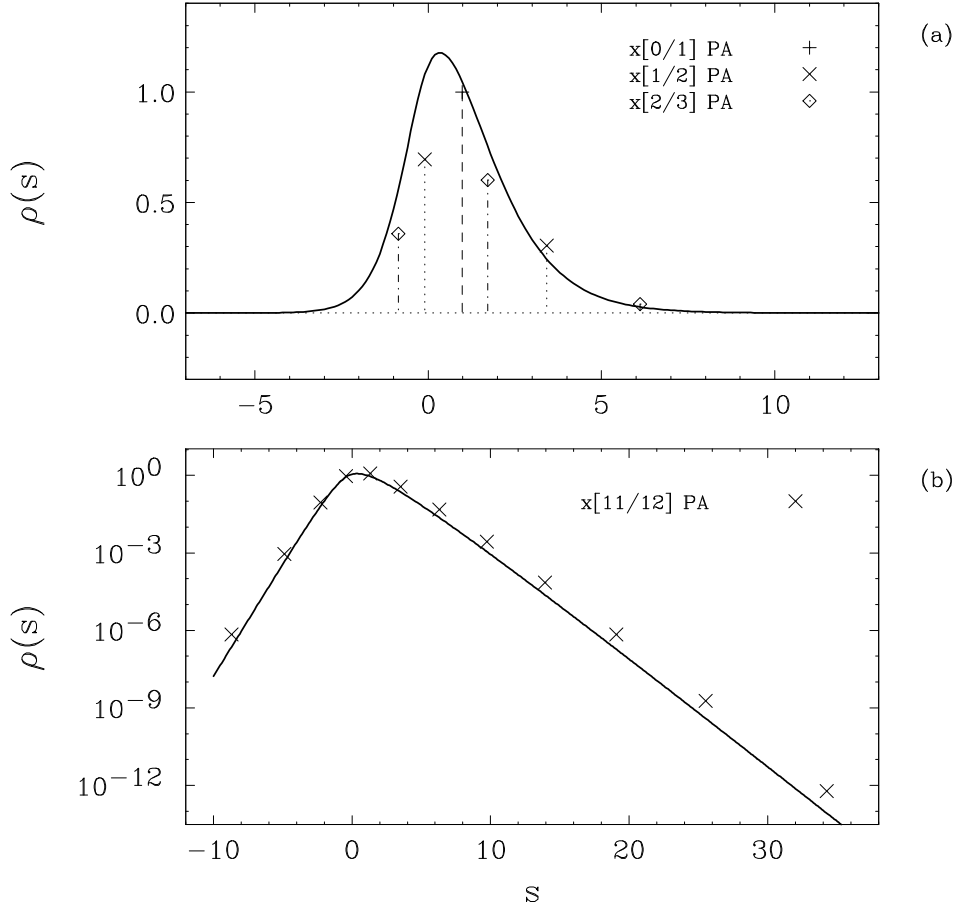,width=12.0truecm,angle=90}
}
\end{center}
\caption{
The momentum distribution function $\rho(s)$ 
in the large-$\beta_0$ limit,
for a virtual gluon in the vacuum-polarization $D$ function. 
In each panel, the 
solid line is the exact result for $\rho(s)$, 
and the different symbols correspond to the locations and
relative strengths (weights) of Dirac $\delta$ functions, as 
determined by diagonal
PA's. As discussed in the text, the $x[N-1/N]$ PA chooses the locations
and weights such that the $2N$
first moments of any function integrated with respect to $\rho(s)$ are 
reproduced exactly. Panel (a) shows low-order PA's on a linear scale, 
and panel (b) shows a representative high-order PA on a logarithmic
vertical scale and an expanded horizontal scale. The
convergence of the PA's to the true momentum distribution function is
clearly visible.}
\label{D_delta}
\end{figure}\newpage

\begin{figure}[htb]
\begin{center}
\mbox{\kern-0.5cm
\epsfig{file=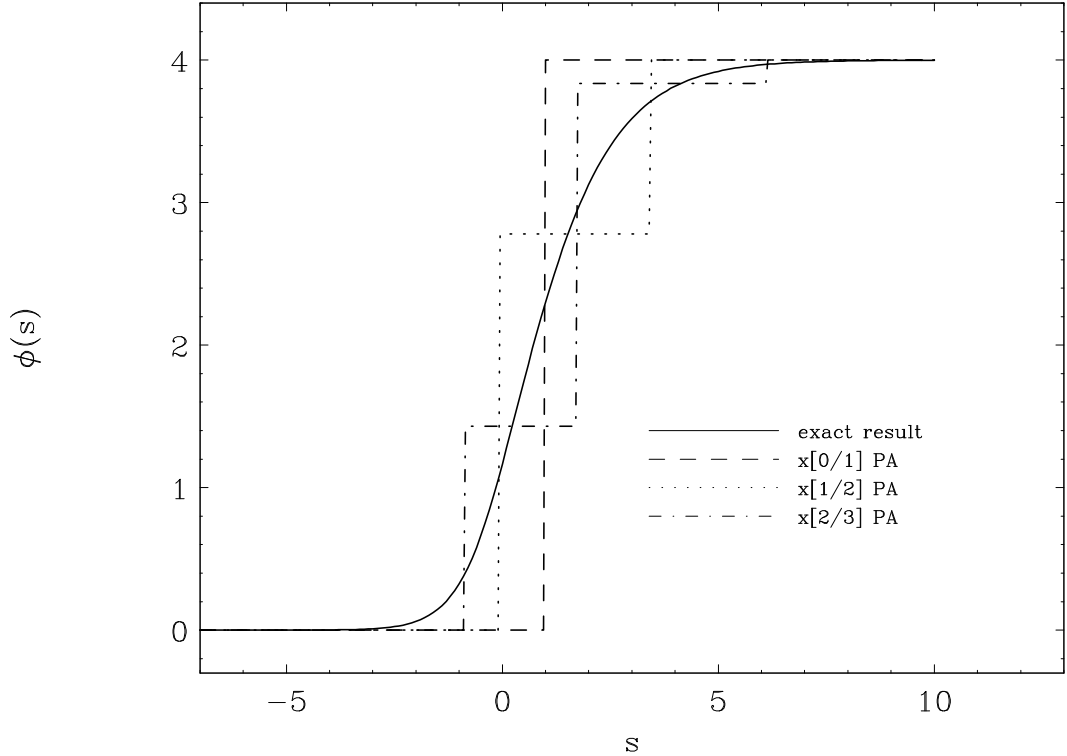,width=10.0truecm,angle=90}
}
\end{center}
\caption{The integral of the momentum distribution function in the large 
$\beta_0$ limit,
for a virtual gluon in the vacuum-polarization $D$ function.
The continuous line represents the exact result for $\phi_D(s)$, and
the other lines describe different approximations to $\phi_D(s)$ that 
correspond to the $x[N-1/N]$ PA's of the perturbative series for 
$N=1,2,3$. The $N$th
approximation to $\phi_D(s)$ is a piecewise-constant function 
composed of $N$ steps, with heights determined by the weights 
of the $\delta$ functions provided by the corresponding PA's.}
\label{D_theta}
\end{figure}

\newpage

\begin{figure}[htb]
\begin{center}
\mbox{\kern-0.5cm
\epsfig{file=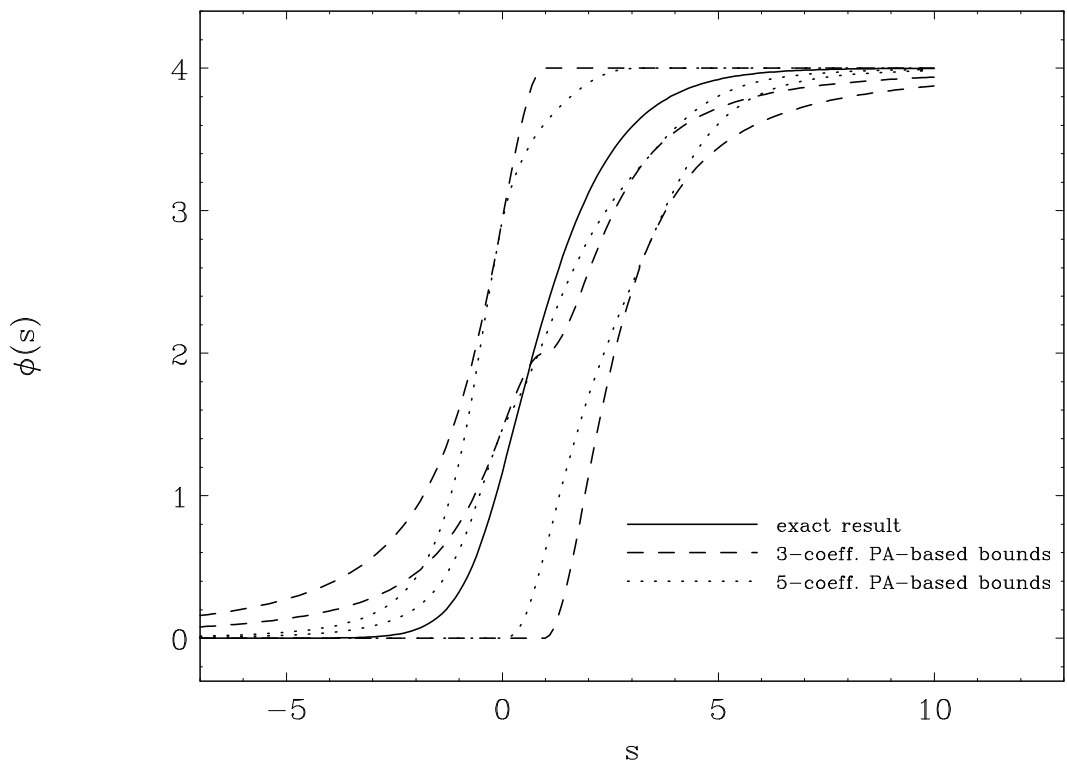,width=10.0truecm,angle=90}
}
\end{center}
\caption{The integral momentum distribution function in the large 
$\beta_0$ limit,
for a virtual gluon in the vacuum-polarization $D$ function.
The continuous line represents the exact result for $\phi_D(s)$, and
the other lines describe the upper and lower bounds, as well as their
averages, that one can construct as described in section
3.5 from the first three (dashed lines) and five (dotted lines)
coefficients of the perturbative series. Note that the average is very
close to the exact distribution function.}
\label{D_bound}
\end{figure}

\newpage
\begin{figure}[htb]
\begin{center}
\mbox{\kern-0.5cm
\epsfig{file=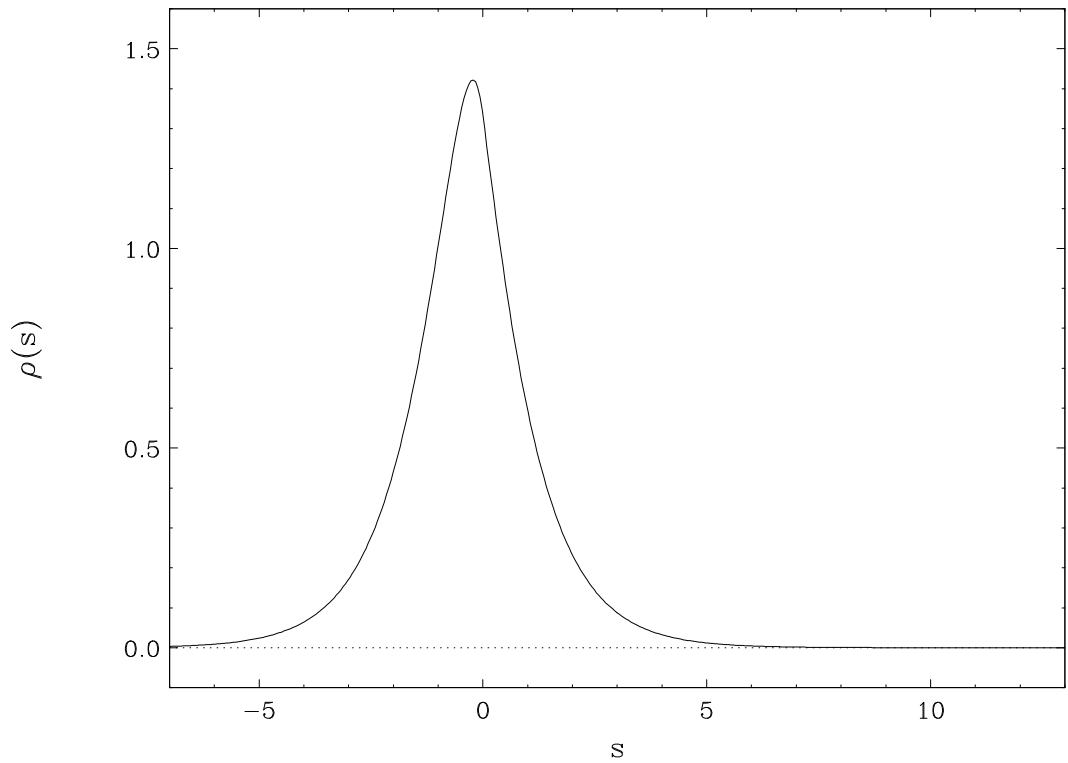,width=10.0truecm,angle=90}
}
\end{center}
\caption{The momentum distribution function $\rho(s)$ 
in the large-$\beta_0$ limit, for a virtual gluon in the Bjorken sum rule.}
\label{D_BjSR}
\end{figure}

\end{document}